\documentclass[iop]{emulateapj}
 \usepackage{psfrag}
 \usepackage{rotating}
 
\shorttitle{Rotation, convective core overshooting, and period changes in classical Cepheids}
\shortauthors{C.L.~Miller, et al.}

\begin{document}

\title{Rotation, convective core overshooting, and period changes in classical Cepheid stellar evolution models}

\author{Cassandra L. Miller\altaffilmark{1,2}}
 \author{Hilding R.~Neilson\altaffilmark{1}}
 \author{Nancy Remage Evans\altaffilmark{3}}
 \author{Scott G. Engle\altaffilmark{4}}
 \author{Edward Guinan\altaffilmark{4}}
  \altaffiltext{1}{David A.~Dunlap Department of Astronomy and Astrophysics, University of Toronto, 50 St. George Street, Toronto, ON, Canada M5S3H4}
  \altaffiltext{2}{Department of Physics and Astronomy, University of British Columbia, 6224 Agricultural Road, Vancouver, BC, Canada V6T 1Z1}
  \altaffiltext{3}{Smithsonian Astrophysical Observatory, MS 4, 60 Garden Street, Cambridge, MA 02138, USA}
  \altaffiltext{4}{Department of Astrophysics \& Planetary Science, Villanova University, 800 Lancaster Avenue, Villanova, PA 19085, USA}

\date{}

\begin{abstract}
Classical Cepheids are  powerful probes of both stellar evolution and near-field cosmology thanks to their great luminosities, pulsations, and that they follow the Leavitt (Period-Luminosity) Law.  However, there still exist a number of questions regarding their evolution, such as the roles of rotation, convective core overshooting and winds. ln particular, how do these processes impact Cepheid evolution and the predicted fundamental properties such as stellar mass.  In this work, we compare a sample of Cepheids with measured rates of period change with new evolution models to test the impact of these first two processes.  In our previous study we found that enhanced mass loss is crucial for describing the sample, and here we continue that analysis but for rotational mixing and core overshooting.  We show that, while rotation is important for stellar evolution studies, rotation, itself, is insufficient to model the distribution of period change rates from the observed sample.  On the other hand,  convective core overshooting is needed to explain the magnitude of the rates of period change, but does not explain the number of stars with positive and negative period change rates. In conclusion, we determine that convective core overshooting and stellar rotation alone are not enough to account for the observed distribution of Cepheid rates of period change and another mechanism, such as  pulsation-driven mass loss, may be required.
\end{abstract}


\section{Introduction}
Cepheid variable stars are  important probes of stellar and cosmological astrophysics thanks to the Leavitt Law \citep{Leavitt1908}. Not only are they essential tools for determining extragalactic distances,  they are also crucial probes of stellar evolution theories.  Cepheid pulsation periods are correlated to the mean density, hence changes in the mean density due to evolution yield changes in pulsation periods \citep{Eddington1918, Szabados1983, Turner2006, Neilson2012a, Neilson2016a}. This relation allows for the direct measurement of stellar evolution and to test state-of-the-art models of stellar evolution.

\cite{Turner2006} measured rates of period change for 196 galactic Cepheids and showed that the rate of period change indicates which crossing of the instability strip the Cepheid is on, which renders them useful tools for studying evolution. They showed that these rates of period change appear consistent with predictions from stellar evolution models.  However, \cite{Neilson2012b,Neilson2012a} went further and found that predictions of period change are inconsistent with that measured for the nearest Cepheid Polaris and that in general stellar evolution theory appears inconsistent with observations.  Classical Cepheids evolve across the instability strip three times: the first is soon after the end of the main sequence as the star expands and cools while the second and third crossing occur when the star transitions from having most energy generated by hydrogen shell burning to helium core burning and the end helium core burning.  During the first and third crossings the pulsation period increases while during the second crossing the period decreases.  By comparing the observed number of stars with positive period change and those with negative period change we have a test of stellar evolution.  \cite{Neilson2012a} computed population synthesis models from a grid of stellar evolution tracks and found that from theory about 85\% of Cepheids should have positive period change.  The fraction from the \cite{Turner2006} sample is about 67\%.  The predicted fraction decreased to about 70\% if the stellar models underwent enhanced mass loss during the Cepheid stages of evolution, suggesting evidence of Cepheid mass loss \citep{Kervella2006, Marengo2010, Matthews2012}.

On the other hand, \cite{Anderson2014} computed stellar evolution models of Cepheids using the Geneva code that included rotation  and moderate convective core overshooting.  They showed that when stars are born with about 50\% of critical rotation the resulting Cepheid blue loop was a different shape and is more luminous than for a stellar evolution model with the same initial mass but no rotation.  In terms of period change, they found that the rotating models have predicted rates of period change consistent with the results from \cite{Turner2006}. Furthermore, \cite{Anderson2014} argued that rotation could help resolve the Cepheid mass discrepancy \citep{Bono2006}. 

The Cepheid mass discrepancy is the difference between masses of Cepheids measured from stellar evolution modelling and from stellar pulsation measurements \citep{Cox1980}.  
\cite{Caputo2005}, \cite{Keller2006} and \cite{Keller2008} showed that the Cepheid mass discrepancy is about  10 - 20\%.  As noted by \cite{Anderson2014,Anderson2016}, rotation is one possibility to solve the mass discrepancy, but \cite{Bono2006} suggested stellar mass loss, convective core overshooting and missing opacities as well.  The potential for missing opacities was deemed unlikely by \cite{Bono2006}.   Period change offers a powerful test of the role of these physical processes.

Convective core overshooting is a phenomenon that leads to a more massive stellar core at the end of a star's main sequence lifetime, thereby changing the mass-luminosity relation of Cepheids.  In main sequence stars with convective cores, convective eddies are assumed to rise towards the top of the convection zone with some acceleration and velocity. At the top of the convection zone the acceleration goes to zero but the eddy velocity does not.  However, the layers above this boundary are not convectively unstable according to the Schwarzschild criteria, therefore in  models the eddy does not penetrate the layers above the convective boundary.  This is unphysical and to account for the fact that the eddies should rise above the convective boundary stellar evolution models include overshooting. This allows convective eddies to penetrate some distance above the core and mix material back into the core. In the \cite{Yoon2005} code we use in this work, overshooting is input by the user as a fraction of the pressure scale height. This overshooting acts to extend the main sequence lifetime and create a more massive post-main sequence core. Because of this convective core overshooting is a possible solution to the mass discrepancy problem as it leads to a more massive helium core \citep{Huang1983} in the progenitor of the Cepheid, which will cause the Cepheid to be more luminous.

As such convective core overshooting could resolve the Cepheid mass discrepancy to a point.  If the discrepancy were the same for all Cepheids then overshooting would be a likely solution.  However, \cite{Keller2008} found significant variation of the mass discrepancy for Galactic Cepheids suggesting that overshooting on its own is insufficient. On the other hand, \cite{Neilson2011} determined that the combination of pulsation-driven mass loss and  convective core overshooting could resolve the discrepancy as measured by \cite{Keller2008}. 

  Because there are a number of processes that can resolve the Cepheid mass discrepancy, we need alternative methods and observables to test those processes. In this work, we return to the analysis of \cite{Neilson2012b} to compare population synthesis models of Cepheid stellar evolution with rotation included to determine how rotation impacts predicted rates of period change.  In the next section, we discuss the stellar evolution model grid along with the included physics of rotation and overshooting in the models.  In Sect.~\ref{pop-syn}, we describe the population synthesis modeling using our stellar evolution tracks. We conclude with a discussion around the impact of our results and the role of stellar rotation in understanding the evolution of classical Cepheids.

\section{Stellar evolution models}
Stellar evolution models were computed using the Binary Evolution Code (BEC), a 1-D hydrodynamical code \cite{Yoon2005} designed to model the evolution of single and binary stars. Stellar evolution models were computed assuming a solar-like metallicity, i.e., $Z = 0.02$, and using \cite{Grevesse1998} opacity tables.  The initial masses of the stars ranged from 3 - 15 M$_\odot$ in steps of 1 M$_\odot$.  The helium abundance in the models is $Y = 0.28$.  Varying the initial helium abundance could impact the evolution of the star and predicted Cepheid pulsation properties, especially the blue and red boundaries of the instability strip \citep{Marconi2009}.  The goal of this work is to consider the importance of rotation, and we will consider variations of the helium abundance in future work. The amount of initial rotational velocity and convective core overshooting parameter varied among the models, with initial rotation ranging from 0 to 350 km/s in steps of 25 km/s.  Convective core overshooting is defined in the code as $\alpha_c H_{P}$. $H_{P}$ is the local pressure scale height at the canonical boundary of the convective core and $\alpha_c$ is a free parameter input by the user.  Over the main sequence lifetime of the star, material will mix a distance of  $\alpha_c H_{P}$ above the convective boundary.  In our analysis we vary the convective core overshooting parameter ranging from $\alpha_c$ = 0.0 to 0.3 in steps of 0.1.  

Mass loss is treated identically to that of \cite{Neilson2012b}, in which the \cite{deJager1988} formulation for estimating $\dot{M}$ is used for cool stars and the \cite{Kudritzki1989} mass-loss rates are used for hot stars.  However, there is no special nor specific prescription assumed for  mass loss during the Cepheid stage of evolution.  We considered arbitrary prescriptions for mass loss in previous works \cite{Neilson2011, Neilson2012b, Neilson2012a} and found that mass loss can explain both the mass discrepancy and the measured distribution of Cepheid rates of period change. However, none of these prescriptions are ideal.  For instance, \cite{Neilson2008} presented a model of pulsation-driven mass loss that is caused by shock propagation in the photosphere.  That model is highly non-linear, in that small changes in parameters describing the wind could result in orders-of-magnitude changes in predicted mass-loss rates. Furthermore, it is not clear that this model is accurate enough to test in stellar evolution codes beyond predicting mass-loss rates of the order $10^{-7}$ -- $10^{-6}~M_\odot$/yr \citep{Neilson2011}.  It is also unclear how mass loss in Cepheids should depend on rotation.  Therefore, for this work we focus on rotation only.

 \begin{figure}[t]
   \begin{center}
   \plotone{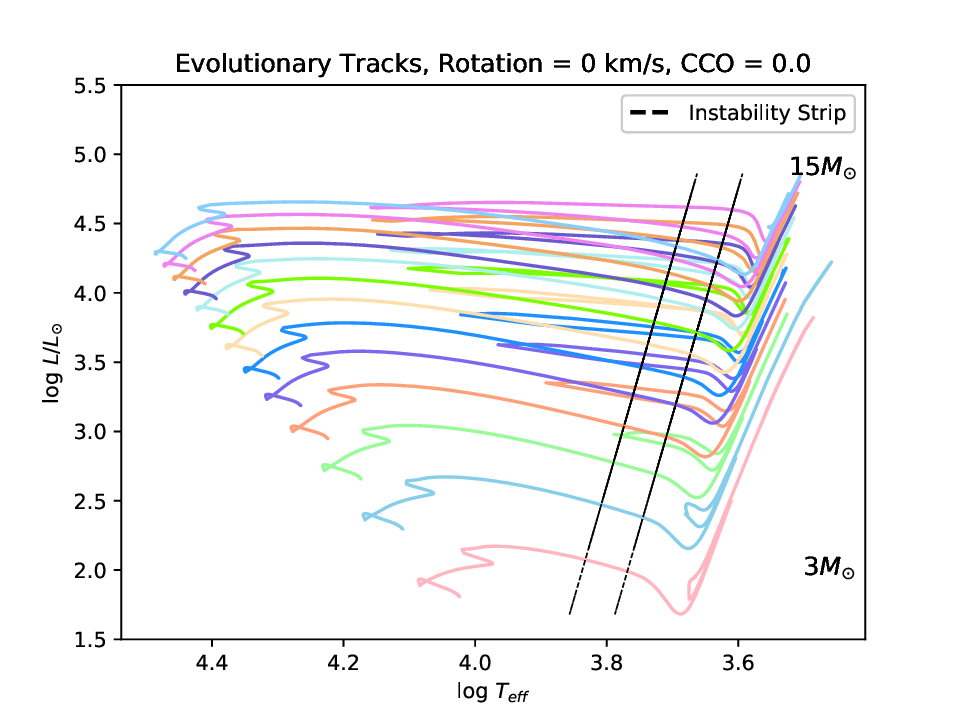}
     \plotone{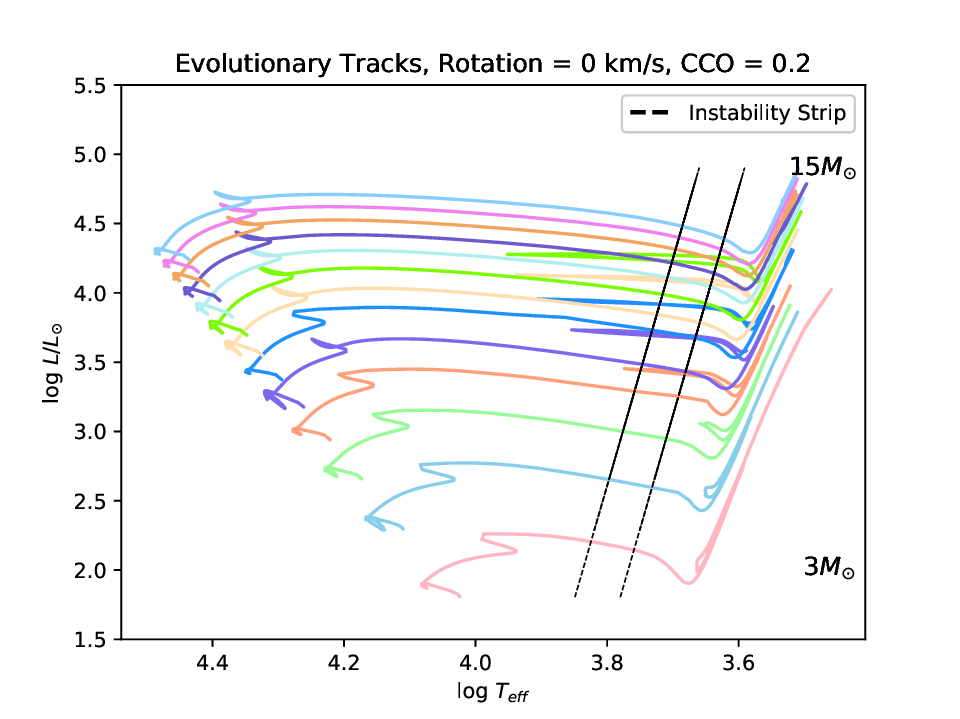}
     \end{center}
      \caption{(Top) Evolutionary Tracks for 3-15 $M_\odot$ stars, in steps of 1~$M_\odot$, with no rotational component and no convective core overshooting ($\alpha_c = 0$). The instability strip is seen in black. (Bottom) Stellar evolution tracks with moderate convective core overshooting, $\alpha_c = 0.2$ with zero rotation for the same mass range. }
         \label{0cco0_evotrack}
   \end{figure}

\begin{figure}
   \centering
   \plotone{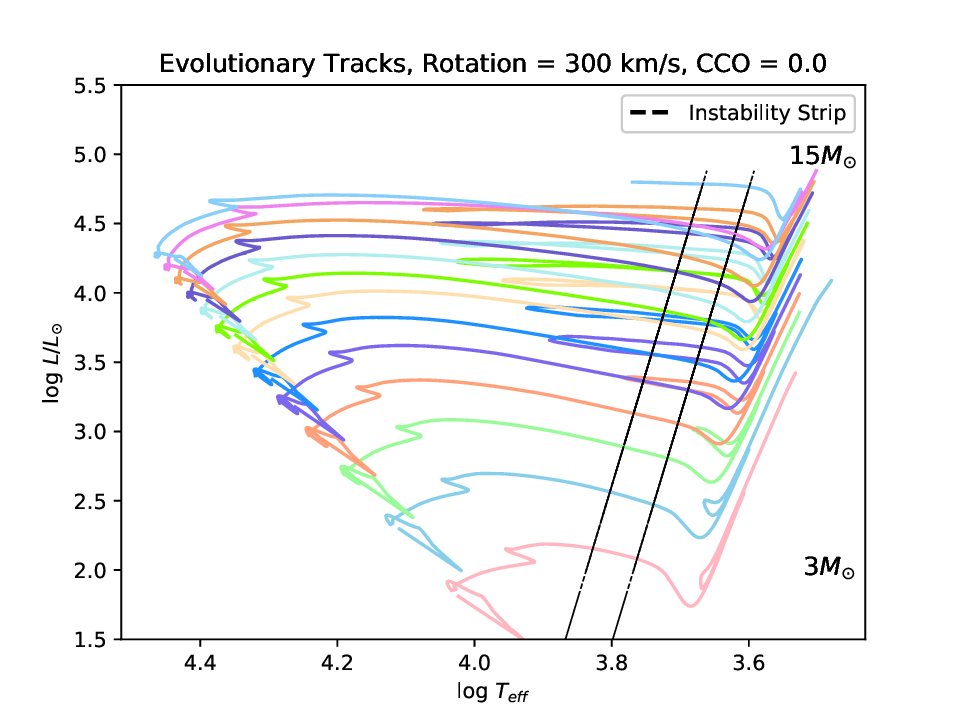}
    \plotone{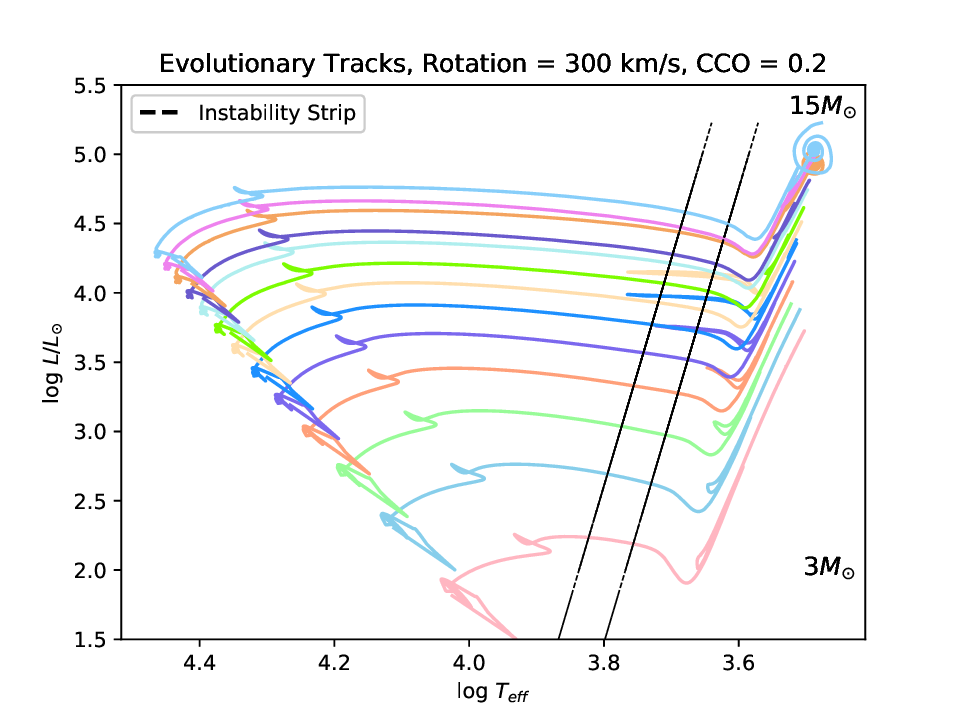}
      \caption{(Top) Evolutionary Tracks for 3-15 M$_\odot$ stars with an iniitial rotation rate $v_{rot} = 300$~km~s$^{-1}$  and no convective core overshooting ($\alpha_c = 0$). The instability strip is seen in black. (Bottom) Stellar evolution tracks with moderate convective core overshooting, $\alpha_c = 0.2$ and rapid rotation for the same mass range.}
         \label{300cco0_evotrack}
   \end{figure}

Stellar evolution tracks for models created with no rotation and no convective core overshooting are shown in Figure~\ref{0cco0_evotrack}, along with models with no rotation and moderate convective core overshooting. The bounds of the instability strip were determined using a relationship between the luminosity and effective temperature of the blue and red strip described by \cite{Bono2000} for Z = 0.02.  Including moderate convective core overshooting in the models has two effects in regards to the stellar evolution tracks.  The first effect is that the lowest mass blue loops that cross into the instability strip increases with increasing overshooting.  The second effect that the maximum mass for blue evolution decreases with increasing overshooting.  We show in Figure~\ref{300cco0_evotrack} a similar grid of stellar evolution models, but with initial rotational velocity of $300~$km~s$^{-1}$. The evolution tracks for the rapidly rotating stars show similar blue loop behaviour as found by \cite{Anderson2014}.  However, the models with both rapid rotation and moderate convective core overshooting have small blue loops with only three models crossing the Cepheid instability strip.

 We compute the models in steps of $\Delta v = 25$~km~s$^{-1}$ from initial velocities ranging from $0$ to $350$~km~s$^{-1}$. We do this so that we can compute population synthesis models that require understanding the rotational distribution of stars.  We consider distributions measured by \cite{Simondiaz2014} for this work.  However, some of the stellar models will have critical rotational velocities that can be less than some of the rotation rates assumed in the models.  When we compute these models using the BEC code \citep{Yoon2005}, if the rotation rate is greater than critical, the model relaxes to a sub-critical value before evolving. This would mean that we might be overestimating the impact of rapidly rotating stellar models in our population synthesis models. However, because rapidly rotating stars are only a small fraction of the total stellar population, then that overestimate should not be significant. An alternative method is to compute models scaled in terms of critical velocity \citep[e.g.,][]{Anderson2016}, however, this is more difficult to compare models this way statistically.

The rates of period change for the Cepheid models computed with the stellar evolution code are determined using the following relation derived by \cite{Turner2006}:
\begin{equation}
\label{pchange}
\frac{\dot{P}}{P} = \frac{6}{7}\frac{\dot{L}}{L} -  \frac{24}{7}\frac{\dot{T}_{\rm{eff}}}{T_{\rm{eff}}}.
\end{equation}
Equation \ref{pchange} is derived by differentiating the Period-Mean Density relation and assuming that the pulsation constant, $Q$, varies as $P^{1/8}$ \citep{Turner2002}. By using $L$ and $T_{\rm{eff}}$ predicted from the models, and computing $\dot{L}$ and $\dot{T}_{\rm{eff}}$, we can use Equation \ref{pchange} to compute  relative rates of period change. This relation also assumes that the mass-loss rate is small, hence the rate of mass change of the star is much smaller than the rates of change of luminosity and effective temperature.  The Period-Mean Density relation  is equivalent to predictions from linear models.  But, because we assume $Q$ is a function of period then the relation deviates from linear adiabatic relations.  For the prescriptions used in this work this assumption is reasonable.  For instance, \cite{Bono2000} computed non-linear pulsation models of Cepheids and fit a period-luminosity-effective temperature relation.  When we take the time derivative of that relation we find $\dot{P}/{P} \approx 0.67 \dot{L}/L - 3.30\dot{T}_{\rm{eff}}/T_{\rm{eff}}$.  This relation is very similar to the analytic one derived here. It should also be noted that the \cite{Bono2000} relation is computed from models assuming a specific mass-luminosity relation, hence the relation depends on the physics assumed in the models. The non-linear models do not deviate significantly from the period-mean density relation or the empirical version we use here \citep{Bono1999, Bono2000, Marconi2005, Marconi2009}.

\begin{figure}
   \centering
   \plotone{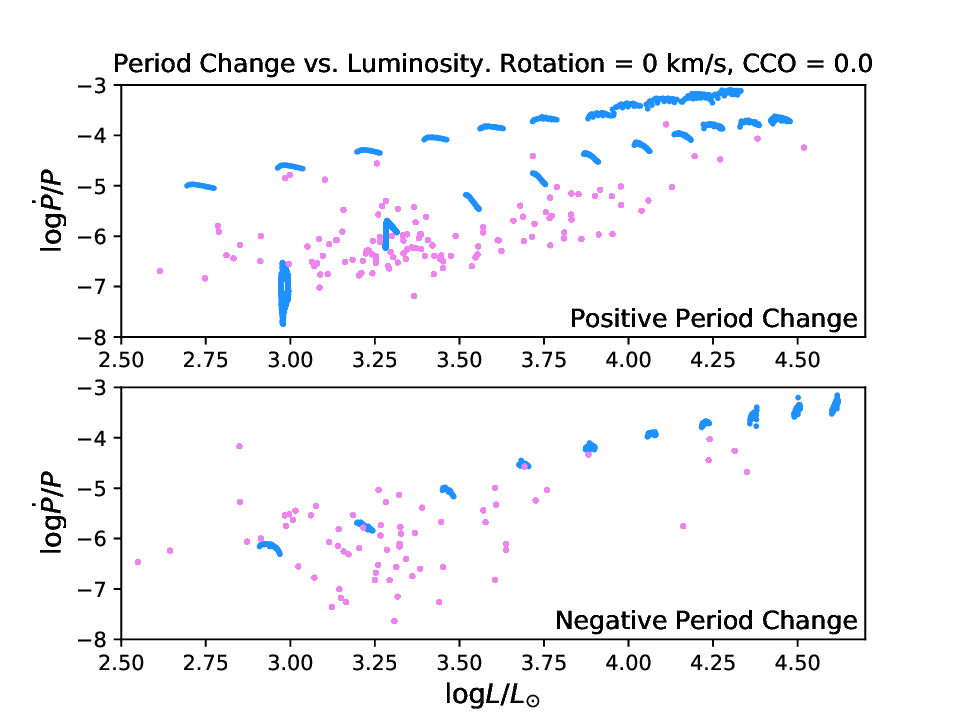}
      \caption{Relative rate of period change as a function of luminosity for 3-15 M$_\odot$ stars with no rotational component and no convective core overshooting ($\alpha_c = 0$) as represented by blue dots. The purple points represent the measured rates from \cite{Turner2006}. The top plot represents Cepheids with a positive rate of period change.  The bottom plot represents Cepheids with a negative rate of period change.   }
         \label{pchange0cco0}
   \end{figure}

We  compute the rates of period change for stars that fall within the Cepheid instability strip of the evolutionary tracks created with our models. A plot of the period change versus the luminosity for models computed with no rotation and no convective core overshooting can be found in Fig.~\ref{pchange0cco0}. Cepheids with a positive rate of period change are separated from those with a negative rate of period change because the rate of the change of the Cepheid's pulsation period is closely tied to it's evolution and mass. A positive rate indicates redward evolution on the first or third crossing of the instability strip, while a negative rate of period change indicates that the Cepheid is on the second crossing.   The rates shown in Fig.~\ref{pchange0cco0} show that comparing predicted rates of period change with those measured can be problematic.  Depending on the Cepheid, the measured rate was built by compiling observations from numerous sources with unclear uncertainties in both timing and brightness. Because of these issues, we consider statistical ratios instead.

 We note that \cite{Turner2006} showed that predicted rates of period change are consistent with observations.  This is somewhat confirmed by \cite{Neilson2012b} in that most predicted rates are consistent with observations, but that it is unclear whether any one specific measurement from observations can be tested.  As such, we continue to consider the comparison in terms of a statistical test.  The issue of specific rates can be seen in greater detail if we consider measurements from other sources. \cite{Neilson2012b} presented a measurement of period change for Polaris that disagrees with that by \cite{Turner2006}. The measurement is difficult because it requires understanding observations spanning more than a century using different techniques from optical telescopes to visual measurement. 
\cite{Piet2001} presented measurements of a sample of Large Magellanic Cloud Cepheids and showed that the measurements are consistent with the \cite{Bono2000} model predications, but not the predictions by \cite{Alibert1999}. \cite{Piet2003} repeated the analysis for a sample of Galactic Cepheids and found discrepancies between model predictions and measurements for the long-period Cepheids. More recently, \cite{Siro2017} measured the rate of period change for the Cepheid VZ~Cyg based on a century of observations, yet noted that the variations were not purely secular and that there were more complex variations as well such as an apparent cyclic variability of about 26~years.  Furthermore, there is an apparent period variability seen in most Cepheids.  \cite{Suveges2018} conducted a detailed statistical analysis of a sample of Cepheid lightcurves and noted an underlying period variability of about 10 - 15~minutes. This is consistent with the period jitter measured by \cite{Derekas2012} and \cite{Evans2015} for a few Cepheids that appears to be a result of convective granulation \citep{Neilson2014b, Derekas2017}.  As a result, there continue to be a number of difficulties in measuring Cepheid period change with high precision and to disentangle secular evolution from apparent period instability, unknown long-term variability, and convection.

\section{Population synthesis modeling}\label{pop-syn}
We compute stellar population models using our grids of stellar evolution tracks.  We assume a \cite{Kroupa2001} initial mass function (IMF) and a constant star formation rate for this analysis.  Because the star formation rate is constant, we can ignore it for the calculations.  Since we are interested in the probability ratio of Cepheids with positive and those with negative rates of period change  the  star formation rate has no impact on the analysis.  The \cite{Kroupa2001} IMF is:
\begin{equation}
\xi(m) \propto m^{-\alpha}
\end{equation}
where $\alpha = 2.3$ for massive stars.  Since we only consider stellar evolution models with masses greater than three solar masses then there is no notable differences with other IMFs \citep[e.g.,][]{Salpeter1955, Chabrier2003}. This prescription is identical to that done previously \citep{Neilson2012b}.

In this work we are considering the role of stellar rotation, hence we need to include the rotational distribution of stars in our calculations.  We use the measured distribution for Galactic OB stars \citep{Simondiaz2014} in the computation.   However, we start by computing the probabilities for each initial rotation rate separately, that is, we assume the entire population is born with a given rotation rate, for instance it might be $0~$km~s$^{-1}$ or 300~km~s$^{-1}$. This is, of course, unphysical, but by doing the computation for each rotation rate, we can measure how rotation impacts the evolutionary  timescales, hence probabilities.  We  fold in the rotational distribution of stars afterwards.  We show that our results do not change significantly as a function of rotation.  As such the choice of rotational distribution will have no consequence.

The population synthesis models  for each initial rotation rate and each assumed amount of convective core overshooting are created for the Cepheids evolved with the BEC. We then determine the probability of a Cepheid having a positive or negative rate of period change and separate the Cepheids into corresponding bins. It is possible to determine the fraction of Cepheids with positive or negative rates of period change by summing the Cepheids in a given bin.   Because we also assume no special prescription for Cepheid mass loss, comparing the computed rates of period change with the observed from \cite{Turner2006} will allow us to determine if a special prescription for mass loss may be necessary to obtain observed results as done by \cite{Neilson2012b}. This is because the rate of period change is directly related to the mass-loss rate $\dot{P}/P \propto \dot{M}/M$ from taking the derivative of the period-mean density relation, however typically $\dot{M}/M << |\dot{P}/P|$. The greater the mass-loss rate the greater and more positive the rate of period change and if the average mass-loss rate is high enough it could cause more Cepheids to have positive rates of period change even while evolving on the second crossing of the instability strip where the rate of period change is expected to be less than zero.

\section{Results}
\begin{table*}[t]
\caption{Predicted fractions of Cepheids with positive and negative period changes}
\begin{center}
\begin{tabular}{ccccc}
\hline
\hline
Convective core overshooting & \multicolumn{2}{c}{$\alpha_c =  0.0 $}&   \multicolumn{2}{c}{$\alpha_c =  0.2$} \\
\hline
Rotation Rate (km~s$^{-1}$) & \% Positive  & \% Negative &  \% Positive  &  \% Negative  \\
\hline
0 &91.7 & 8.3 & 96.5 & 3.5  \\
25 & 90.8& 9.2 &95.9 &4.1 \\
50 & 86.9 & 13.1 &93.9 & 6.1 \\
75 & 79.6 & 20.4 & 89.4 & 10.6 \\
100 & 77.9 & 22.1 & 98.7 & 1.3\\
125 & 77.2 & 22.8 & 85.2 & 14.8\\
150 & 76.2 & 23.8 & 87.2 & 12.8\\
175 & 78.7 & 21.3 & 86.6 & 13.4\\
200 & 74.2 & 25.8 & 95.3 & 4.7\\
225 & 83.8 & 16.2  & 96.9 & 3.1\\
250 & 82.2 & 17.8 &93.9 & 6.1\\
275 & 90.2 & 9.8 & 94.4 & 5.6\\
300 & 91.4 & 8.6 & 94.2 & 5.8\\
325 & 86.5 & 13.5 & 91.5&  8.5\\
350 & 90.5 & 9.5 &99.5 & 0.5\\
\hline
Observed & \multicolumn{2}{c}{\% Positive = 67}&   \multicolumn{2}{c}{\% Negative = 33} \\ 
\hline
\end{tabular}
\label{t1}
\end{center}
\end{table*}

For each initial rotation rate and convective core overshooting value we compute the relative fraction of models with positive and with negative rates of period change.  We find that regardless of the initial rotation rate or convective core overshooting parameter the fraction of Cepheids with positive period change is about 90\% and 10\% for those with negative period change as can be seen in Table~\ref{t1}.  This suggests that period change, at least statistically, is insensitive to the physics of core overshooting and rotational mixing.

This insensitivity has been noted before for overshooting.  \cite{Neilson2012a} and \cite{Neilson2014} found similar results in that overshooting acts to change the mass-luminosity relation for Cepheids.  That is, in terms of period change, Cepheids with different masses have the same rates of period change for different values of core overshooting.  This is because each Cepheid crossing of the instability strip is depends on physics driving stellar evolution.  During the first crossing the evolutionary time scale is related to the Kelvin-Helmholtz time scale, hence is proportional to the mass of the star.  Evolution along the second crossing of the instability strip occurs as the star  generates energy through mostly hydrogen shell burning and is beginning helium-core burning.  The time scale for blueward evolution is driven by the timescale for the star to have shell burning, but not have ignited  significant helium core fusion.  The third crossing is determined by helium core burning, hence its time scale is also a function of helium core mass.  Over the mass range we consider in this paper, overshooting will not change the time scales, hence fraction, of Cepheids on the first crossing, while the impact on the second and third crossings also remain unchanged.

The same argument is also true for rotation.  Rotational mixing impacts the mass of the helium core and rotation can change the shape of the Cepheid blue loop.  However, neither appears to impact the relative time scales of evolution significantly.  The fraction of Cepheids with positive period change decreases to about 77\% for models with zero convective core overshooting.  The fraction is smaller, but not enough to be consistent with the observed fraction of 67\%.   This relative insensitivity of period change to rotation was also seen in the \cite{Anderson2014, Anderson2016}.  Because the fraction of Cepheids with positive and negative rates of period change does not vary as a function of period change then we do not need to worry about the initial rotational distribution of stars in our model.  The result that models predict too many Cepheids with positive rates of period change will not change.  

However, while the model predictions will not change, we check if the observed ratio may be biased for Cepheids with small absolute values of period change.  Some of the measurements from \cite{Turner2006} sample can have large uncertainties particularly when there is only small changes measured in the O-C diagrams \citep[see][]{Neilson2016a}, but we have assumed that any measurement error would be symmetric about zero period change.  We test this assumption by removing all Cepheids with period changes $|\dot{P}| \le 0.1$, $1$, and $10~$s/yr and recompute the observed ratios from the \cite{Turner2006} sample.   From these constraints, the observed fraction of Cepheids with positive period change is $68\%, 61\%$, and $65\%$, respectively.  Based on this quick check, we conclude that the data is essentially not biased in a way such that the actual number ratio is much higher.

Another potential test is to remove Cepheids that can be classified as first-overtone pulsators.  It is not clear that first-overtone pulsators will follow the same distribution as the fundamental-mode pulsators.  One of us (N.R.~Evans) compiled a list of potential first-overtone pulsators from the \cite{Turner2006} data. From the sample, we suggest that 22 Cepheids out of the 196 in the sample are likely first-overtone pulsators.  Of those 22, nine have negative rates of period change and thirteen positive rates.  Once these are removed from the observed sample we find that 67\% of the remaining have positive rates of period change as opposed to 68\% for the entire sample. This difference is not statistically significant and suggests that the combination of different pulsation modes in the data set leads to no bias.

 We conduct a third test where we compute the ratio of positive and negative rates of period change by shifting the boundaries of the instability strip.  We do this by shifting the blue and red edges by 0.1 dex in effective temperature.  When the instability strip is shifted to hotter effective temperatures, the fraction of Cepheids with predicted positive rates of period change increased to an average of about 90 - 95\%.  On the other hand when we shift the instability strip to cooler effective temperatures the fraction of Cepheid with positive rates of period change decreases.  For stars with significant initial rotation the fraction is about 60 - 65\% which would be consistent with observations.  However, in this work the red edge is already significantly cooler than the edge predicted by \cite{Anderson2016}, even before we shift it to cooler temperatures.  As such it is unlikely that the difference between predicted period changes and observed period changes can be understood in terms of the location of the instability strip.
 
\section{Period change around the median}
\begin{figure*}
   \centering
    \plottwo{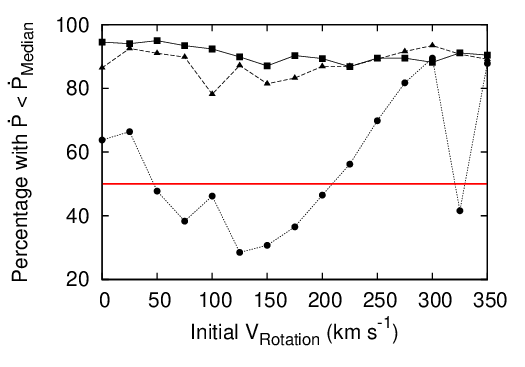}{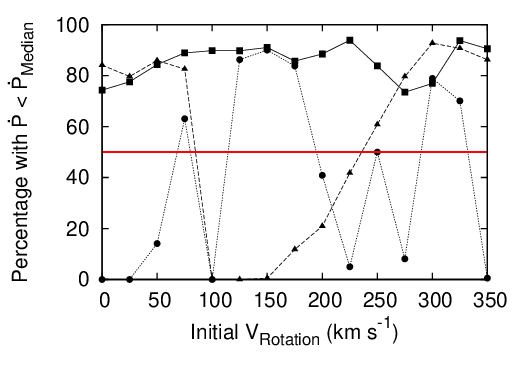}
      \caption{The percentage of Cepheids predicted from stellar evolution models as a function of initial rotational velocity that have rates of period change \emph{less} than the median values from the \cite{Turner2006} sample.  The lines with square points represent models with $\alpha_c = 0$, triangles $\alpha_c = 0.1$, and circles $\alpha_c = 0.2$.  The left plot shows the results for models with positive rates of period change and the right plot those with negative rates of period change.
      }
         \label{hist}
   \end{figure*}

Given this data, we also conduct another population synthesis test.  We compute the median positive and negative rate of period change from the observed data.  We then check from our population synthesis models the fractions of Cepheid with positive and negative rates of period change that are less than the median value from the observations.  This is not a strong statistical test, but it does offer insights into physical processes. We compute this as a function of initial rotation for different convective core overshooting values of $\alpha_c = 0.0, 0.1$, and $0.2$, and present the histograms in Fig.~\ref{hist} along with showing the data for $\alpha_c = 0.0$ and $0.2$ in Table~\ref{t2}.

For the case of zero convective core overshooting we find that most Cepheid models have positive rates of period change less than the median value from the observations, about 90\% for all initial rotation rates, with a slight decrease as a function of rotation.  We find a similar result for Cepheids with negative rates of period change, ranging from about 73\% to 92\%.  The variation of the rates of positive period change can be interpreted as rotational mixing increasing the size of the helium core, hence decreasing the helium burning lifetime during the third crossing of the Cepheid instability strip. That is, a Cepheid with a more massive helium core will behave like a more massive star without rotational mixing. On the other hand the data also suggest that rotation and rotational mixing tends to slow the evolution on the instability strip, at least until an initial rotation rate of 225~km/s is assumed.  Based on this result, we can conclude that models with zero overshooting are inconsistent with this simple comparison to the median of the observed data such that models suggest that most Cepheids would have rates of positive and negative period change less than the median value from the observations.

On the other hand, when we consider stellar evolution models with moderate convective core overshooting, $\alpha_c = 0.2$, the population changes significantly. On average the population of Cepheids with positive period change greater than the observed median is greater than that for models with no convective core overshooting. For the grid with zero initial rotation, the fraction is 36\%, increasing to 71\% for a grid with initial rotation of 125~km/s.  We apply a rotation distribution from \cite{Simondiaz2014} to compute the weighted fraction of Cepheids with positive period change in this case and find that about 51\% of Cepheids are predicted to have rates of positive period change greater than the median observed value. This is opposed to nearly 85\% of Cepheid models assuming zero overshooting. 

The large changes in the fractions from one initial rotation rate to the next relate to changes in the blue loop structures of stellar models.  For a given rotation rate and convective core overshooting parameter, blue loops can evolve to an effective temperature hotter than the blue edge of the instability strip where a Cepheid would spend most of its time.  In other words Cepheids with small negative period change tend to be near the tip of the blue loop.  If the star evolves beyond the blue edge of the instability strip then there will be few models with small negative period change ($\dot{P} > \dot{P}_{\rm{Median}}$).  If the tip of the blue loop is just cooler than the blue edge than we will find many models with small negative rates of period change.  Because we are computing a distribution assuming an IMF then the probability will be dominated by smallest mass models that form blue loops.  Therefore, the large shifts in the probability are due to small changes in blue loop structure as a function of mass and initial rotation. The changes are more prevalent for models with greater overshooting because the blue loops for those models begin to cross the instability strip at a greater minimum stellar mass.

\begin{table*}[t]
\caption{Predicted fractions of the rate of period change in Cepheids less than the observed median value }
\begin{center}
\begin{tabular}{ccccc}
\hline
\hline
Convective core overshooting & \multicolumn{2}{c}{$\alpha_c =  0.0 $}&   \multicolumn{2}{c}{$\alpha_c =  0.2$} \\
\hline
Rotation Rate (km~s$^{-1}$) & \%  $<$ Positive $\dot{P}/P$ & \% Negative $<$  $\dot{P}/P$&  \% $<$ Positive $\dot{P}/P$  &  \%  $<$Negative  $\dot{P}/P$  \\
\hline
0 & 94.5& 74.3 & 63.8 & 0\\
25 & 94.0& 77.3 & 66.4& 0\\
50 &95.0 & 84.3 & 47.8& 14.4\\
75 &93.4 & 88.9 & 38.3& 63.0\\
100 &92.4& 89.8 & 46.2& 0\\
125 &89.9& 89.8& 28.5& 86.3\\
150 &87.1& 91.0 & 30.7& 90.0\\
175 &90.3& 85.7 & 36.5&83.7\\
200 &89.3& 88.5 & 46.5 & 40.9\\
225 &86.8& 93.9 & 56.2& 5.0\\
250 &89.5 & 83.8& 69.8& 50.0\\
275 &89.5& 73.5 & 81.7& 8.1\\
300 &88.2& 76.9& 89.5 & 78.9\\
325 &91.1& 93.8 & 41.6& 70.1\\
350 &90.4 &90.5 & 87.8 & 0.5\\
\hline

\end{tabular}
\label{t2}
\end{center}
\end{table*}

For Cepheids with negative period change the fraction with rates of period change greater than the median is about 67\% when we include rotation and moderate convective core overshooting in the analysis.  If no overshooting is considered then the fraction is about 15\%.  Again, the result suggests that moderate overshooting is necessary for modelling Cepheid evolution. Rotation is also an important ingredient, but neither solves the problem that evolutionary models predict too many stars with positive period change.

\section{Discussion}
In this work we have tested our understanding of Cepheid evolution and period change using population synthesis modeling as well as comparing the results with period change measurements from \cite{Turner2006}.  In the models we include tests of convective core overshooting and rotation to complement our earlier work on stellar mass loss \citep{Neilson2012b}.  We continue to find that overshooting does not change the relative fraction Cepheids with positive and negative period change.  Rotation also does not impact this ratio. 

We also compared our population synthesis models and predicted what fraction of the models have period changes greater and less than the median value from the observations for both negative and positive rates of period change. In this situation we do find that the amount of convective core overshooting is important such that models with moderate convective core overshooting are more consistent with observations than no overshooting.  We emphasize that this is not a fit, only that moderate overshooting appears more consistent.  Furthermore, rotation does not appear to have a significant impact, but is an important ingredient.

We did not test enhanced mass loss in this work. We showed previously enhanced mass loss can reduce the fraction of Cepheids with positive period change and is consistent with the observed fraction \citep{Neilson2012b}. We also did not test how enhanced mass loss impacts the fractions of Cepheids with period change greater than or less than the observed median rates of period change. This is because we do not yet have a theory for pulsation-enhanced mass loss that can provide a test at the level of necessary detail.  \cite{Neilson2008,Neilson2009} developed an analytic mass-loss prescription based on pulsation-driven shocks propagating in the photosphere.  This theory provided estimates of mass-loss rates consistent with infrared interferometric observations \citep[e.g.,][]{Kervella2006, Merand2006, Gallenne2013} and is marginally consistent with infrared and radio observations of the prototype $\delta$~Cephei \citep{Marengo2010, Matthews2012}.  The theory was also incorporated into stellar evolution models to show that enhanced mass loss plus convective core overshooting can explain the Cepheid mass discrepancy \citep{Keller2008, Neilson2011}. However, this theory requires knowledge of pulsation properties and amplitudes and is not ideal for predicting mass-loss rates in Cepheids with precision, hence is not a reasonable test of period change ratios around the median observation.  

\cite{Neilson2012b} tested the impact of mass loss on Cepheid evolution and period change for a constant mass-loss rate.  That impact is negligible for $\dot{M}/M_\ast << |\dot{P}/P|$. If the mass-loss rate is large enough to impact the period change then some Cepheids with positive period change will shift from being less than the median value to being greater than the median.  Furthermore, some Cepheids with negative rates of period change will shift to small positive rates of period change.  \cite{Neilson2012b} found that the fraction of Cepheids with negative period changes increases when mass-loss rates are high enough, suggesting that enhanced mass loss increases the timescale of the second crossing  and changes the period change probabilities.  That is more Cepheids will have large, negative rates of period change and is shown in Fig.~3 of that paper.  Assuming a constant mass-loss rate appears consistent with this scenario, however, there is no precision enhanced mass-loss model that can directly test this.

These results offer hints for our understanding of the Cepheid mass discrepancy \citep{Keller2008}. \cite{Bono2006} offered four potential resolutions to this long-standing problem: missing opacities, enhanced mass loss, rotational mixing, and convective core overshooting.  \cite{Bono2006} then demonstrates that unknown missing opacities cannot account for more than small fraction of that discrepancy given the success of the current opacities for modeling stellar evolution of other stars, leaving three options.  Previous works have demonstrated that each of the three processes are important for understanding Cepheid evolution \citep{Huang1983, Neilson2008, Neilson2009, Anderson2016}, but it is becoming clear that each ingredient is important in different ways.  For instance, convective core overshooting appears to be crucial for understanding the evolution of the Cepheid eclipsing binary  OGLE-LMC-CEP0227 \citep{Piet2010, Cassisi2011, Neilson2012c, Prada2012}. Mass loss appears important for understanding infrared observations of Cepheids, whereas rotation helps understand anomalous abundances in some Cepheids \citep{Anderson2014, Neilson2014}.

We conclude that rotation does not impact the period change distribution and hence we require additional physics to explain the measured distribution, thereby, strengthening the results of \cite{Neilson2012b}. With respect to the Cepheid mass discrepancy, we conclude that we must consider all three ingredients (rotation, mass loss, and convective core overshooting) to understand Cepheid evolution. Moderate convective core overshooting and rotation must be considered for models to be consistent with the evolution of mass sequence progenitors and enhanced mass loss is required to account for infrared observations.  The combination will produce a population of models consistent with the distribution of Cepheid period changes.

\begin{acknowledgements}
C.L.M. and H.R.N acknowledge that most of this research was conducted at the University of Toronto. The University of Toronto operates on traditional land of the Huron-Wendat, the Seneca, and most recently, the Mississaugas of the Credit River and we are grateful to have the opportunity to work on this land. N.R.E is grateful for support provided from the Chandra X-ray Center NASA 
Contract NAS8-03060. E.F.G and S.G.E. are grateful for grants from NASA to Villanova University.

\end{acknowledgements}

\bibliographystyle{aa}
\bibliography{pdot-1}

\end{document}